# Measuring the Magnetic Flux Density with Flux Loops and Hall Probes in the CMS Magnet Flux Return Yoke

B. Curé, N. Amapane, A. Ball, A. Gaddi, H. Gerwig, A. Hervé, V. I. Klyukhin, R. Loveless, and M. Mulders

*Abstract*—The Compact Muon Solenoid (CMS) is a general purpose detector, designed to run at the highest luminosity at the CERN Large Hadron Collider (LHC). Its distinctive features include a 4 T superconducting solenoid with 6-m-diameter by 12.5-m-length free bore, enclosed inside a 10,000-ton return yoke made of construction steel. The flux return yoke consists of five dodecagonal three-layered barrel wheels and four end-cap disks at each end comprised of steel blocks up to 620 mm thick, which serve as the absorber plates of the muon detection system. To measure the field in and around the steel, a system of 22 flux loops and 82 3-D Hall sensors is installed on the return yoke blocks. A TOSCA 3-D model of the CMS magnet is developed to describe the magnetic field everywhere outside the tracking volume that was measured with the field-mapping machine. The voltages induced in the flux loops by the magnetic flux changing during the CMS magnet standard ramps down are measured with six 16-bit DAQ modules. The off-line integration of the induced voltages reconstructs the magnetic flux density in the yoke steel blocks at the operational magnet current of 18.164 kA. The results of the flux loop measurements during three magnet ramps down are presented and discussed.

## I. Introduction

THE magnetic flux density in the central part of the Compact Muon Solenoid (CMS) detector [1], [2], where the tracker and electromagnetic calorimeter are located, was measured with precision of $7 \cdot 10^{-4}$ with the field-mapping machine at five central field values of 2, 3, 3.5, 3.8, and 4 T [3]. To describe the magnetic flux everywhere outside this measured volume, a three-dimensional (3-D) magnetic field model of the CMS magnet has been developed [4] and calculated with the program TOSCA [5] from Cobham CTS Limited. The model reproduces the magnetic flux density distribution measured with the field-mapping machine inside the CMS coil within 0.1% [6]. During the Large Hadron Collider (LHC) long shutdown occurred in 2013/2014 the CMS magnet flux return yoke was upgraded with two additional 14 m diameter end-cap disks at the extremes of the muon detection system. The CMS magnet model was modified and validated by comparing the calculated magnetic flux density with the measured one in the selected regions of the CMS magnetic system. [7].

A direct measurement of the magnetic flux density in the yoke selected regions was performed during the CMS magnet test of 2006 with 22 flux loops of 315–450 turns of the 45-wire flat ribbon cable wound around the yoke blocks in special grooves of 30 mm wide and 12–13 mm deep. The cross-sections of areas enclosed by the flux loops vary from 0.3 to 1.59 m² on the yoke barrel wheels and from 0.5 to 1.12 m² on the yoke end-cap disks. The total areas covered by the flux loops are calculated on the basis of each individual wire turn and vary from 122 to 642 m². In 2006, the "fast" discharges of the CMS coil (190 s time-constant) made possible by the protection system, which is provided to protect the magnet in the event of major faults, induced in the flux loops the voltages caused by the magnetic flux changes. An integration technique [8] was developed to reconstruct the average initial magnetic flux density in steel blocks at the full magnet excitation. The contribution of the eddy currents to the magnetic flux was calculated with the program ELECTRA [9] from Cobham CTS Limited and estimated on the level of a few per cent [10].

The comparisons of measurements done with the flux loops and the CMS magnet model values calculated with the program TOSCA are described elsewhere [11].

During the LHC long shutdown of 2013/2014 the read out system of the flux loop voltages was upgraded and includes now six DAQ modules USB-1608G of Measurement Computing with 8 differential 16-bit analog inputs each. The DAQ modules are attached by the USB cables to two network-enabled AnywhereUSB®/5 hubs connected to the DAQ PC through 3Com® OfficeConnect® Dual Speed Switch 5 and optical fiber cable of 100 m with two Magnum CS14H-12VDC Convertor Switches. This modification allows to perform the flux loops measurements during the CMS magnet standard ramps down with the current discharge speed as low as 1–1.5 A/s with acceptable accuracy.

## II. COMPARISON OF THE MEASURED AND CALCULATED MAGNETIC FLUX DENSITY

The measurements used for comparisons are obtained in three CMS magnet standard discharges from the current of 18.164 kA to zero done in 2015 and 2016 as shown in Fig. 1.

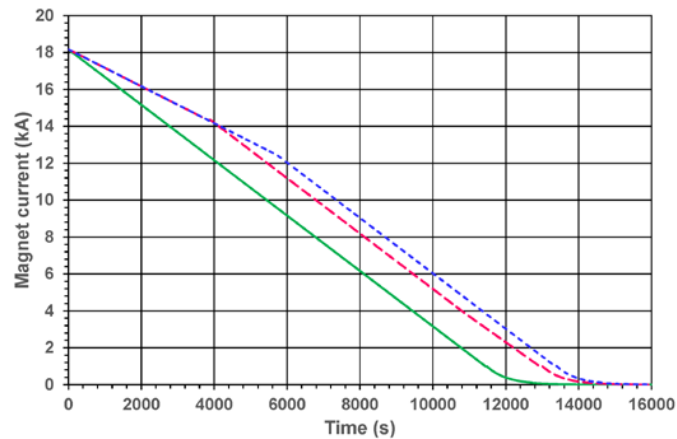

Fig. 1. CMS magnet current standard discharges from 18.164 kA to zero made on July 17–18, 2015 (smooth line), September 21–22, 2015 (dashed line), and September 10, 2016 (dotted line).

The first discharge of July 17–18, 2015 was made with the constant current ramp down speed of 1.5 A/s to the current of 1 kA, and after stop of 42 s the fast dump of the magnet was triggered manually. The measurements of the voltages with maximum amplitudes of 20–250 mV induced in the flux loops are integrated during 15061.5 s in the flux loops located on the barrel wheels and during 15561.5 s in the flux loops located on the end-cap disks.

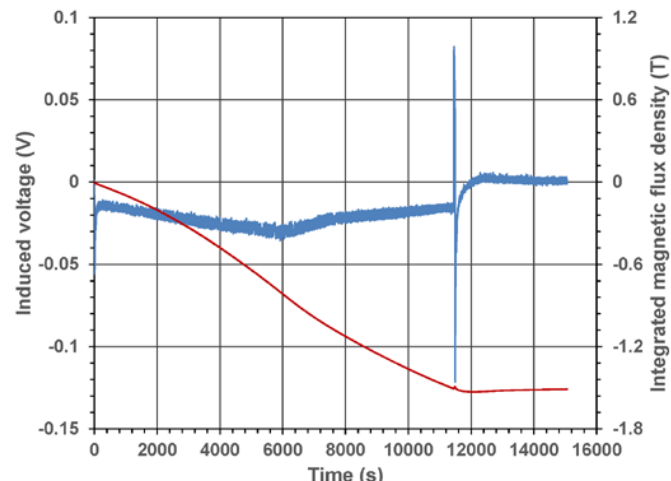

Fig. 2. The induced voltage (left scale, noisy curve) and the integrated average magnetic flux density (right scale, smooth curve) in the cross section at $Z=0$ m of the first layer block of the central barrel wheel. The rapid maximum and minimum voltages at 11,445 s correspond to the short stop in the magnet charge down at the current of 1 kA, and the subsequent transition from the standard discharge to the fast discharge of the magnet.

The typical induced voltages in the first magnet ramp down together with the integrated average magnetic flux densities are shown in Figs. 2 and 3. The rapid maximum and minimum voltage at 11,445 s corresponds to the stop of the ramp down at the current of 1 kA for 42 s and the following transition from the standard ramp down to the fast discharge of the magnet on the external resistor.

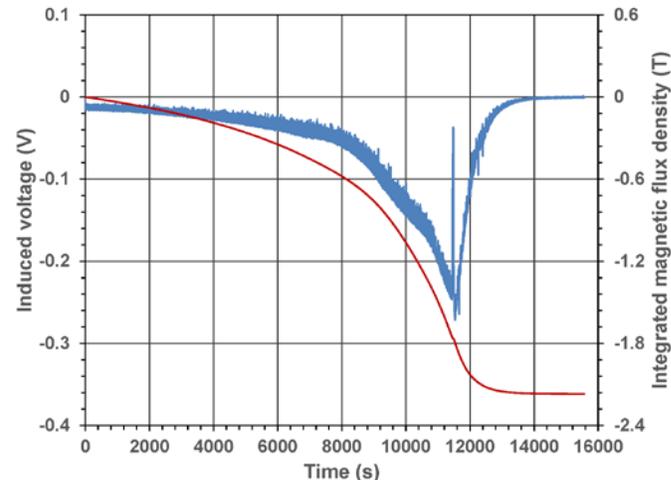

Fig. 3. The induced voltage (left scale, noisy curve) and the integrated average magnetic flux density (right scale, smooth curve) in the cross section at $Y=-4.565$ m of the first end-cap disk block. The rapid maximum and minimum voltages at 11,445 s correspond to the short stop in the magnet charge down at the current of 1 kA, and the subsequent transition from the standard discharge to the fast discharge of the magnet.

The coordinate system used in this study corresponds to the CMS reference system where the $X$-axis is directed in the horizontal plane toward the LHC center, the $Y$-axis is directed upward, and the $Z$-axis coincides with the superconducting coil axis and has the same direction as the positive axial component of the magnetic flux density.

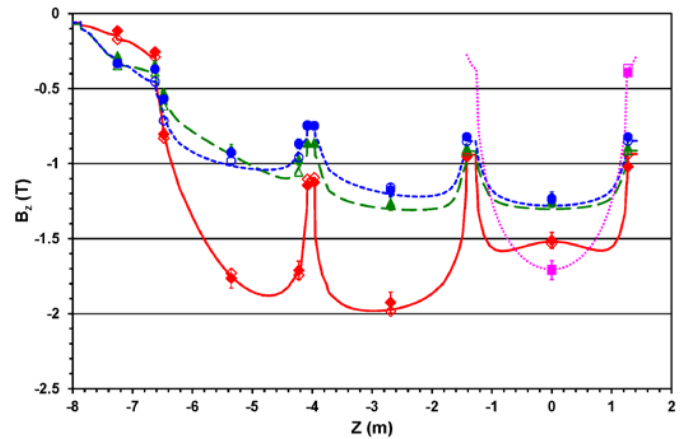

Fig. 4. Axial magnetic flux density measured (filled markers) and calculated (open markers) in the tail catcher (squares) and the first (rhombs), second (triangles), and third (circles) barrel layers vs. the $Z$-coordinate. The lines represent the calculated values along the Hall sensor locations at near side of the yoke and at the $Y$-coordinates of $-3.958$ m (small dotted line), $-4.805$ m (solid line), $-5.66$ m (dashed line), and $-6.685$ m (dotted line).

The second magnet discharge of September 21–22, 2015 was performed with two constant ramp down speeds: 1 A/s to the current of 14340 A, and 1.5 A/s to the current of 1 kA. The third magnet discharge of September 10, 2016 was similar, but the current at a transition from 1 A/s to 1.5 A/s speed was 12480 A. In both these magnet ramps down the fast discharges were triggered from the current of 1 kA, and the off-line integration of the induced voltages was performed during 17,000 s.

The flux loops measurements are complemented with measuring the magnetic flux density with the 3-D Hall sensors installed between the barrel wheels and on the end-cap disks at the axial Z-coordinates of 1.273, –1.418, –3.964, –4.079, –6.625, and –7.251 m. The sensors are aligned in the rows at the vertical Y-coordinates of –3.958, –4.805, –5.66, and –6.685 [7] on two sides of the magnet yoke: near to the LHC center (positive X-coordinates), and far from the LHC center (negative X-coordinates).

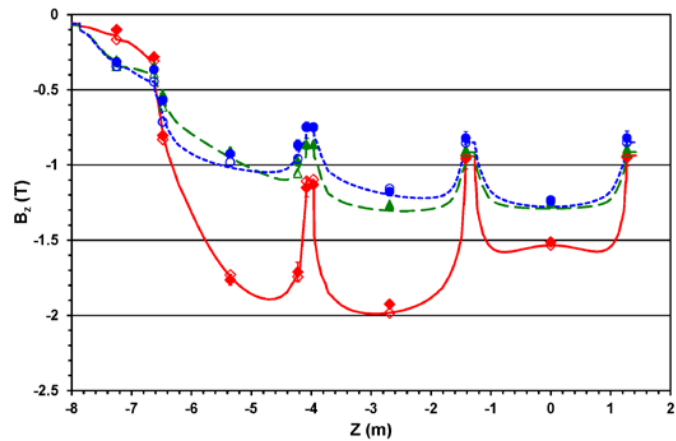

Fig. 5. Axial magnetic flux density measured (filled markers) and calculated (open markers) in the first (rhombs), second (triangles), and third (circles) barrel layers vs. the Z-coordinate. The lines represent the calculated values along the Hall sensor locations at far side of the yoke and at the Y-coordinates of −3.958 m (small dotted line), −4.805 m (solid line), −5.66 m (dashed line), and −6.685 m (dotted line).

In Figs. 4–6 the measured values of the magnetic flux density vs. Z- and Y-coordinates are displayed and compared with the calculated field values obtained with the CMS model at the operational current of 18.164 kA.

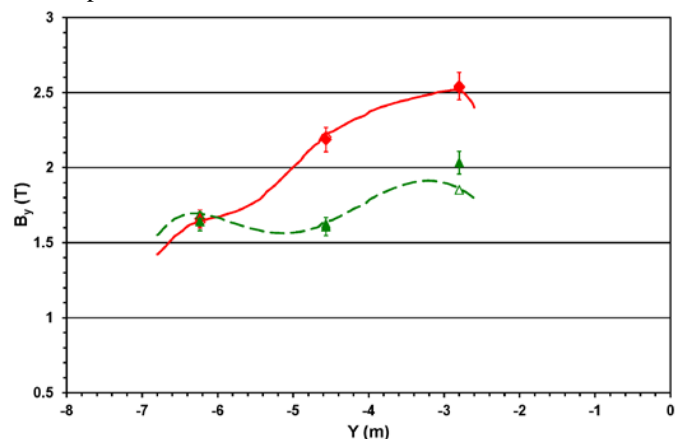

Fig. 6. Axial magnetic flux density measured (filled markers) and calculated (open markers) in the first (rhombs) and second (triangles) end-cap disks vs. the Y-coordinate. The lines represent the calculated values in the middle planes of the end-cap disks.

The comparison gives the differences between the calculated and measured values of the magnetic flux density in the flux loop cross-sections as follows: 4.71 ± 7.38 % in the barrel wheels and −1.03 ± 4.12 % in the end-cap disks. The error bars of the magnetic flux density measured with the flux loops include the standard deviation in the set of three measurements (9.3 ± 6.3 mT or 0.71 ± 0.55 % in average) and the systematic error of ±3.59 % aroused from the flux loop conductor arrangement. The difference between the calculated and measured magnetic flux density in the 3-D Hall sensor locations is (4 ± 8) %. The error bars of the Hall sensor measurements are ± (0.011±0.003) mT.

In 2006 the comparison gave the differences between the calculated and measured values of the magnetic flux density as follows [7]: 0.59 ± 7.41 % in the barrel wheels and −4.05 ± 1.97 % in the end-cap disks at the maximum current of 17.55 kA; 1.41 ± 7.15 % in the barrel wheels and −2.87 ± 2.00 % in the end-cap disks at the maximum current of 19.14 kA.

## III. CONCLUSIONS

The measurement of the magnetic flux density in the steel blocks of the CMS magnet flux return yoke is made using the flux loop technique and the standards magnet discharge with the current ramp down speed of 1–1.5 A/s.

The precision of the measurements is at the same level compared to the results obtained in 2006 using the fast discharge of the magnet with the time constant of 190 s.